\documentclass[letterpaper]{article} % DO NOT CHANGE THIS
\usepackage{aaai24}  % DO NOT CHANGE THIS
\usepackage{times}  % DO NOT CHANGE THIS
\usepackage{helvet}  % DO NOT CHANGE THIS
\usepackage{courier}  % DO NOT CHANGE THIS
\usepackage[hyphens]{url}  % DO NOT CHANGE THIS
\usepackage{graphicx} % DO NOT CHANGE THIS
\usepackage{natbib}  % DO NOT CHANGE THIS AND DO NOT ADD ANY OPTIONS TO IT
\usepackage{caption} % DO NOT CHANGE THIS AND DO NOT ADD ANY OPTIONS TO IT
\usepackage{algorithm}
\usepackage{algorithmic}
\usepackage{amsmath}
\usepackage{float}
\usepackage{multicol}
\usepackage{multirow}
\usepackage{bm}
\usepackage{enumitem}
\usepackage{bbding}
\usepackage{booktabs} % for professional tables
\usepackage{dsfont}
\usepackage{array}
\usepackage{diagbox}

\DeclareMathAlphabet{\pazocal}{OMS}{zplm}{m}{n}

\DeclareMathAlphabet\mathbfcal{OMS}{cmsy}{b}{n}

\usepackage{newfloat}
\usepackage{listings}
\DeclareCaptionStyle{ruled}{labelfont=normalfont,labelsep=colon,strut=off} % DO NOT CHANGE THIS
\floatstyle{ruled}
\newfloat{listing}{tb}{lst}{}
\floatname{listing}{Listing}
\title{SoundCount: Sound Counting from Raw Audio with \\ Dyadic Decomposition Neural Network}
\author{
    Yuhang He\textsuperscript{\rm 1}, Zhuangzhuang Dai\textsuperscript{\rm 2}, Long Chen\textsuperscript{\rm 3,4}\thanks{corresponding author.}, Niki Trigoni\textsuperscript{\rm 1}, Andrew Markham \textsuperscript{\rm 1}\\
}
\affiliations{
    \textsuperscript{\rm 1}Department of Computer Science,
    University of Oxford, UK.
    yuhang.he@cs.ox.ac.uk\\
    \textsuperscript{\rm 2}Department of Applied AI and Robotics, Aston University. UK\\
    \textsuperscript{\rm 3}Institute of Automation, Chinese Academy of Sciences, China.\\
    \textsuperscript{\rm 4}WAYTOUS Ltd., China.\\
}

\usepackage{bibentry}
\begin{document}

\maketitle

\begin{abstract}
In this paper, we study an underexplored, yet important and challenging problem: counting the number of distinct sounds in raw audio characterized by a high degree of polyphonicity. We do so by systematically proposing a novel end-to-end trainable neural network~(which we call DyDecNet, consisting of a dyadic decomposition front-end and backbone network), and quantifying the difficulty level of counting depending on sound polyphonicity. The dyadic decomposition front-end progressively decomposes the raw waveform dyadically along the frequency axis to obtain time-frequency representation in multi-stage, coarse-to-fine manner. Each intermediate waveform convolved by a parent filter is further processed by a pair of child filters that evenly split the parent filter's carried frequency response, with the higher-half child filter encoding the \textit{detail} and lower-half child filter encoding the \textit{approximation}. We further introduce an energy gain normalization to normalize sound loudness variance and spectrum overlap, and apply it to each intermediate parent waveform before feeding it to the two child filters. To better quantify sound counting difficulty level, we further design three polyphony-aware metrics: \textit{polyphony ratio}, \textit{max polyphony} and \textit{mean polyphony}. We test DyDecNet on various datasets to show its superiority, and we further show dyadic decomposition network can be used as a general front-end to tackle other acoustic tasks. Code: \url{github.com/yuhanghe01/SoundCount}.
\end{abstract}

\section{Introduction}
Suppose you went to the seaside and heard a cacophony of seagulls, squawking and squabbling. An interesting question that naturally arises is whether you can tell the number of seagulls flocking around you from the sound you heard? Although a trivial example, this sound ``crowd counting'' problem has a number of important applications. For example, passive acoustic monitoring is widely used to record sounds in natural habitats, which provides measures of ecosystem diversity and density~\cite{jmse9070685,sounddensityest,soundscape_dataset}. Sound counting helps to quantify and map sound pollution by counting the number of individual polluting events~\cite{Bello2018}. It can also be used in music content analysis~\cite{openmic_data}. Despite its importance, research on sound counting has far lagged behind than its well-established crowd counting counterparts from either images~\cite{multicol_crowdcount,wang2019learning}, video~\cite{li2018csrnet} or joint audio-visual~\cite{ambientSound}.

We conjecture the lack of exploration stems from three main factors. First, sound counting has long been treated as an over-solved problem by sound event detection~(SED) methods~\cite{sed_tutorial,CRNNNet,seldnet,sounddet}, in which SED goes further to identify each sound event's~(\textit{e.g.} a bird call) start time, end time and semantic identity. Sound counting number then becomes easily accessible by simply adding up all detected events. Secondly, current SED only tags whether a class of sound event is present within a window, regardless of the number of concurrent sound sources of the same class like a series of baby crying or multiple bird calls~\cite{polyphonic_aed_icassp22}. Thirdly, labelling acoustic data is technically-harder and more time-consuming than labelling images, due to the overlap of concurrent and diverse sources. The lack of well-labelled sound data in crowded sound scenes naturally hampers research progress. Existing SED sound datasets~\cite{seldnet,tutsed2009} capture simple acoustic scenarios with low polyphony and where the event variance is small. The simplified acoustic scenario in turn makes sound counting task by SED methods tackleable. But when the sound scene becomes more complex with highly concurrent sound events, SED methods soon lose their capability in discriminating different sound events~\cite{Pankajakshan2019PolyphonicSE,CRNNNet}. In the meantime, some researchers think sound counting is equivalent to sound source separation task~\cite{multaker_ASR,sed_sep,compmen_ssep,subakan2022resourceefficient,twostep_ssep}, in which the sound is counted as the source number by isolating individual sound from sound mixture and assigning it to corresponding sound source. However, our proposed sound counting is different from source number counting, it directly counts the overlapping events number, regardless of if these events come from the same sound source. Therefore, a study specific for sound counting problem is desirable and overdue.

In this paper, we study the general sound counting problem under highly polyphonic, cluttered and concurrent situations. Whilst the challenges of image-based crowd counting mainly lie in spatial density, occlusion and view perspective distortion, the sound counting challenges are two-fold. Firstly, acoustic scenes are additive mixtures of sound along both time and frequency axes, making counting overlapping sounds difficult~(temporal concurrence and spectrum-overlap). Secondly, there is a large variance in event loudness due to spherical signal attenuation with distance.

To tackle these challenges, we propose a novel dyadic decomposition neural network to learn a sound density representation capable of estimating cardinality directly from raw sound waveform. Unlike existing sound waveform processing methods that all apply frequency-selective filters on the raw waveform in single stage~\cite{sounddet,ein_v2,iclr_LEAF,sounddoa,MFCC}, our network progressively decomposes raw sound waveform in a dyadic manner, where the intermediate waveform convolved by each parent filter is further processed by its two child filters. The two child filters evenly split the parent filter's frequency response, with one child filter encoding the waveform \textit{approximation}~(the one with the lower-half frequency response) and the other one encoding the waveform \textit{details}~(the one with the higher-half frequency response). To accommodate sound loudness variance, spectrum-overlap and time-concurrence, we further propose an energy gain normalization module to regularize each intermediate parent waveform before feeding it to two child filters for further processing. This hierarchical dyadic decomposition front-end enables the neural network to learn a robust TF representation in multi-stage coarse-to-fine manner, while introducing negligible extra computation cost. By setting each filter's frequency cutoff parameters to be learnable and self-adjustable during optimization in a data-driven way, the final learned TF representation can better characterize sound existence in time and frequency domain. Following the front-end, we add a backbone network to continue to learn a time framewise representation. Such representation can be used to derive the final sound count number by either directly regressing the count number, regressing density map~(the one we choose) or following SED pipeline. Apart from the network, we further propose three polyphony-aware metrics to quantify sound counting task difficulty level: polyphony ratio, maximum polyphony and mean polyphony. We will give detailed discussion to show the feasibility of three metrics.

We run experiment on large amounts of sound datasets, including commonly heard bioacoustic, indoor and outdoor, real-world and synthetic sound. Comprehensive experimental results show the superiority of our proposed framework in counting under different challenging acoustic scenarios. We further show our proposed dyadic decomposition front-end can be used to tackle other acoustic task, like SELD~\cite{seldnet,sounddet}. In summary, we make three main contributions: \textbf{First}, propose dyadic decomposition front-end to decompose the raw waveform in a multi-stage, coarse-to-fine manner, which better handles loudness variance, spectrum-overlap and time-concurrence. \textbf{Second}, propose a new set of polyphony-aware evaluation metrics to comprehensively and objectively quantify sound counting difficulty level. \textbf{Third}, show DyDecNet superiority on various counting datasets, and its potential to be used as a general learnable TF extraction front-end.

\begin{figure*}[t]
    \centering
    \includegraphics[width=0.90\linewidth]{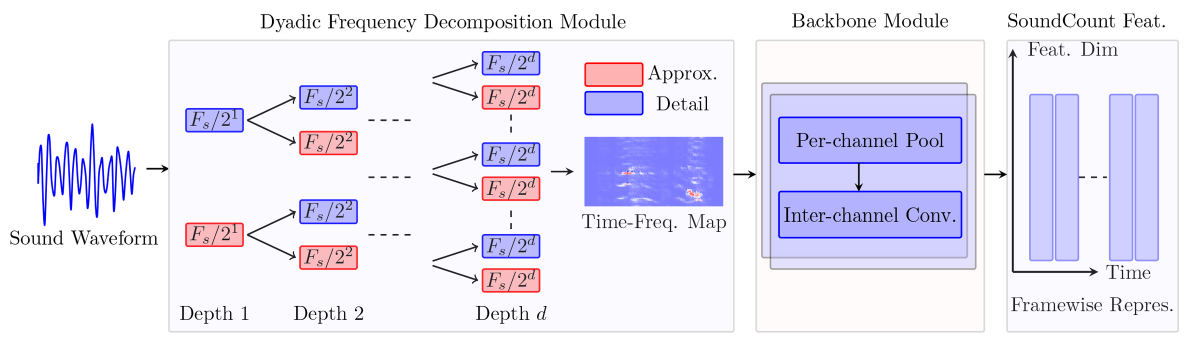}
    \caption{DyDecNet pipeline. We first feed the input raw sound waveform to the dyadic decomposition front-end to learn a time-frequency representation, which is further fed to a backbone neural network to continue to learn framewise representation. Such representation retains time information, so it is general enough to get count number by either regression or SED method. The dyadic decomposition front-end consists of a set of parameterized learnable band-pass filters. Each intermediate waveform processed by a parent filter is further processed by two child filters, with lower-half filter~(red color) encoding \textit{approximation} and higher-half filter~(light-blue) encoding \textit{details}.}
    \label{fig:dyadic_pipeline}
\end{figure*}

\section{Dyadic Decomposition Neural Network}

We put the related work discussion and sound counting task definition in Appendix due the space limit.
Different sound classes typically exhibit different spectral properties. A canonical way to process raw sound waveform is to apply a frequency-selective filter bank $\mathbfcal{F}_f = \{f_i\}_{i=1}^{k}$ to project the raw sound waveform onto different frequency bins. Traditional Fourier transform\,\cite{MFCC} or Wavelet transform\,\cite{mallat_wavelet} construct fixed filter banks in which all filter-construction relevant hyperparameters are empirically chosen and thus may not be optimal for a particular task. Recent methods~\citep{sounddet,iclr_LEAF} relax some hyperparameters to be trainable so that the filter bank can be optimized in a data-driven way. A learnable filter bank often leads to better performance than fixed filters. However, all existing methods apply all filters, either learnable or fixed, on the raw waveform in a one-stage manner. Such shallow and one-stage processing may fail to learn powerful and robust representation for sound counting task where large loudness variance and heavy spectrum overlap exist. In our dyadic decomposition framework, we instead adopt a progressive pairwise decomposition strategy to obtain the time-frequency~(TF) representation. It learns a TF representation from coarse to fine-grained granularity. Particularly, it consists of a dyadic frontend and a backbone.

\subsection{Dyadic Frequency Decomposition Frontend}

In dyadic decomposition frontend, we construct a set of $D$ hierarchical filter banks $\mathbfcal{F}_{dyadic}^D = \{\mathbfcal{F}_{2^1}^1, \mathbfcal{F}_{2^2}^2, \cdots, \mathbfcal{F}_{2^D}^D\}$. The $d$-th filter bank has $2^d$ filters, each filter is parameterized by a learnable high frequency-cutoff parameter and a low frequency-cutoff parameter. By cascading these filter banks, we consecutively decompose the raw waveform in frequency domain dyadically, leading to coarse-grained to fine-grained TF representation. Specifically, we denote the dyadic filter banks depth by $D$, in the depth $d$ filter bank $\mathbfcal{F}_{2^d}^d$, we have $2^d$ filters evenly divide the waveform sampling frequency $F_s$. Therefore, each single filter's frequency response length is $\frac{F_s}{2^{d}}$, the $i$-th filter $f_i^d$ high frequency cutoff $F_h$ and low frequency cutoff $F_l$ are initialized as,

\begin{equation}
    F_h(f_i^d) = \frac{F_s}{2^d}\cdot (i+1), \ \ \ F_l(f_i^d) = \frac{F_s}{2^d}\cdot i
% \end{aligned}
\label{eqn1}
\end{equation}

%where $F(\cdot)$ is the frequency response function for a filter, it evenly divides input waveform's frequency sampling frequency and assigns the each filter with the same amount of frequency range. 
From Eqn.\,(\ref{eqn1}) we can see that dyadic decomposition frontend forms a complete binary-tree-like structure, in which the filter number doubles and each filter's frequency response length halves as the tree's depth increases by one. The intermediate waveform processed by a ``parent'' filter is just further processed by its two ``children'' filters. The frequency responses of the two children filters evenly split their parent filter's frequency response. The child filter carrying the higher half frequency response encode the parent's processed intermediate waveform's \textit{detail} while the other one carrying the lower half frequency response instead encodes the \textit{approximation}. For example, for the filter $f_i^d$ in the $d$-th filter bank, its frequency response lies in $[\frac{F_s}{2^d}\cdot i, \frac{F_s}{2^d}\cdot (i+1)]$, its two children filters $f_{2i}^{d+1}$ and $f_{2i+1}^{d+1}$ in the depth $d+1$ evenly divide its frequency range, so $f_{2i}^{d+1}$ carries $[\frac{F_s}{2^d}\cdot i, \frac{F_s}{2^d}(i+\frac{1}{2})]$. $f_{2i+1}^{d+1}$ carries $[\frac{F_s}{2^d}(i+\frac{1}{2}), \frac{F_s}{2^d}\cdot (i+1)]$.

With the pre-constructed dyadic decomposition filter banks, we cascade them together to process the raw sound waveform, progressively learning the final TF representation. In our implementation, each filter in dyadic filter banks is a learnable band-pass filter. We adopt rectangular band-pass in frequency domain filter which comprises of a learnable high frequency cutoff parameter $F_{h}$ and a learnable low frequency cutoff parameter $F_{l}$. Converting it to time domain through the inverse Fourier transform, we get $sinc(\cdot)$ function like filter that is used to convolve with the waveform. For example, the filter $f_i^d$ in Eqn.\,(\ref{eqn1}) is represented as,

\begin{equation}
    f_i^d[t, F_h, F_l] = 2F_{h}sinc(2\pi F_{h}t) - 2F_{l}sinc(2\pi F_{l}t)
    \label{eqn2}
\end{equation}

\begin{figure*}[t]
    \centering
    \includegraphics[width=0.95\linewidth]{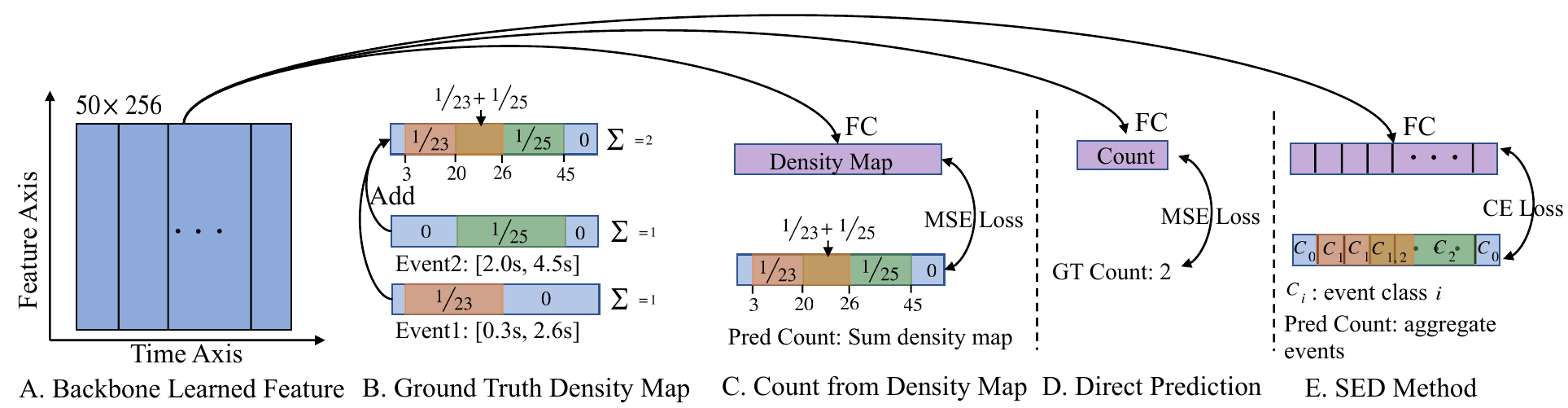}
    \caption{Three counting methods illustration. For density map~(sub-fig.~C), the sum~(or integral) of the density map equals to the count number. We can also direct regress the final count number~(sub-fig.~D), or use SED method~(sub-fig.~E).}
    \label{fig:density_map_illu}
    \vspace{-2mm}
\end{figure*}

where $sinc(x)=sin(x)/x$, $t$ indicates the filter's representation at time $t$. $F_h$ and $F_l$ are initialized according to Eqn.\,(\ref{eqn1}), but they can be further adjusted during the training process. $sinc(\cdot)$ filters have been successfully used in speech recognition~\citep{SincNet} and sound event detection and localization~\citep{sounddet}. In our dyadic decomposition frontend, each filter from different depth has separate and independent learnable parameters\,(high frequency cutoff and low frequency cutoff). Moreover, our constructed filter is much longer\,(1025 in our case) than traditional 1D/2D Conv filters\,(3 or 5). Its wide length characteristic enables the filter to have a wide field-of-view on the raw waveform. Cascading them together allows the filters in later layers\,(larger depth) to have an even wider field-of-view on the input raw waveform. With this advantage, we do not have to model sound event temporal dependency explicitly with RNN network. As a result, the whole dyadic frequency decomposition frontend is fully convolutional and parametrically learnable, it is parameter-frugal and computationally efficient. In practice, the dyadic decomposition frontend depth is 8, so the output TF representation has 256 frequency bins. At the same time, we downsample the intermediate waveform by 2 before feeding it to its two children filters in the initial 5 dyadic filter banks to reduce the memory cost.

\subsection{Energy Gain Normalization}

\begin{figure}[t]
\centering
  \centering
  \includegraphics[width=0.90\linewidth]{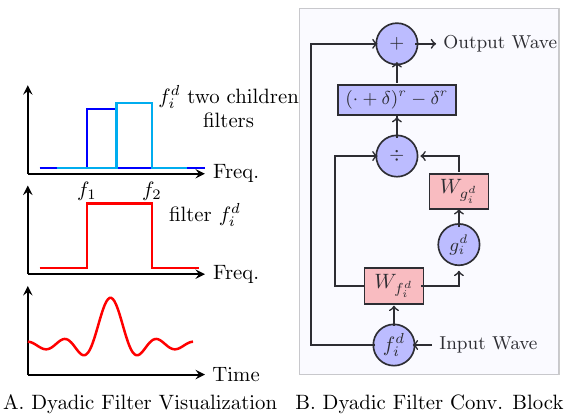}
  \captionof{figure}{\footnotesize Dyadic filter illustration. Left: In time domain, dyadic filter is a $sinc$ function curve. In frequency domain, dyadic filter is a rectangular band-pass filter with learnable high frequency $f_2$ and low frequency cutoff $f_1$. The filter's two child filters\,(left-top) evenly splits the parent filter's frequency response. Right: dyadic filter convolution block. The input waveform is fed to an energy normalization module. Then a skip-connection is added.}
  \label{fig:dyadicfilter_illu}
  \vspace{-4mm}
\end{figure}

We further design an energy gain normalization module to regularize each intermediate waveform before feeding them to the next dyadic filter bank. The motivation of introducing energy gain normalization is two-fold: first, to reduce sound event loudness variance led by sound events' different spatial locations; Second, to reinforce the frontend to learn to better tackle spectrum overlap challenge led by intra-class sound events in the sound scene. Specifically, for the intermediate waveform $W_{f_i^d}$ processed by a dyadic filter $f_i^d$, we first smooth it with a learnable 1D Gaussian kernel $g_i^d$ parameterized by learnable width $\sigma$ to get the corresponding smoothed waveform $W_{g_i^d}$ which just contains loudness. We then introduce a learnable automatic gain control parameter $\alpha$ to mitigate sound loudness impact. Furthermore, another two learnable compression parameters $\delta$ and $\gamma$ are introduced to further compress $W_{f_i^d}$. The overall energy gain normalization can be represented as,

\begin{equation}
    W_{f_i^d} = (\frac{W_{f_i^d}}{(W_{g_i^d})^\alpha} + \delta)^\gamma - \delta^{\gamma}
\label{eqn4}
\end{equation}

where $\alpha$, $\delta$ and $\gamma$ are learnable parameters. As a result, the energy gain normalization \textit{eg}-Norm is fully learnable and parameterized by four learnable parameters \textit{eg}-Norm($\sigma, \alpha, \delta, \gamma$). Practically, each filter in dyadic filter banks is associated with an independent \textit{eg}-Norm module. Similar energy normalization has been successfully used in tasks like keyword spotting~\citep{PCEN,PCEN_howwhy}. The difference lies in the fact that they apply exponential moving average operation to get smoothed waveform representation, so the computation is very slow because it iterates along the time axis to compute the averaged value step by step. Our proposed energy gain normalized strategy instead adopts a Gaussian kernel to get the smoothed waveform, in which it can be easily implemented as 1D convolution. The dyadic filter visualization and energy normalization module is shown in Fig.~\ref{fig:dyadicfilter_illu}.

\subsection{Backbone Neural Network}

We add a lightweight backbone neural network to the frontend neural network to further learn a representation useful for call counting . The backbone network consists of two parts: per-channel pooling and inter-channel 1D convolution. Unlike existing methods~\cite{CRNNNet,seldnet} that first convert 1D sound waveform into 2D map with fixed FFT-like transform, then learn from the 2D map with 2D Conv. operations, our method directly learns from sound raw waveform with learnable 1D Conv.. Specifically, we downsample each channel separately by assigning each channel with an independent frequency-sensitive learnable filter. We call such learnable downsampling per-channel pooling. It helps to learn sound event's frequency variance along the time axis individually. Moreover, we add normal 1D Conv. to achieve inter-channel communication, which enhances the neural network to learn concurrent sound events interaction. The backbone serves as the backend to learn framewise representation for counting.

\subsection{Density Map and Loss Function}

The backbone network discussed above learns a framewise representation $[T_b, F_b]$, where $T_b$ indicates the time steps and $F_b$ indicates feature size. There are three potential ways to derive the final sound count number from the learned representation: 1. directly regress the count number; 2.SED method: detect sound events first and then aggregate results to get final count; 3. predict the density map. For a sound event with time location $[t_1, t_2]$, its density map is a 1D vector with value $\frac{1}{t_2-t_1}$ during its occurrence time, otherwise it is 0. So the count number equals the vector integral. We show the regressing density map produces the best result~(see Table~\ref{table:countmethod}). We thus adopt the mean squared error~(\textbf{MSE}) loss during training to directly regress the density map. The comparison of three methods is shown in Fig.~\ref{fig:density_map_illu}.

\section{Counting Difficulty Quantification}
\label{sec:evaluation}

\begin{figure}[t]
    \centering
  \includegraphics[width=0.90\linewidth]{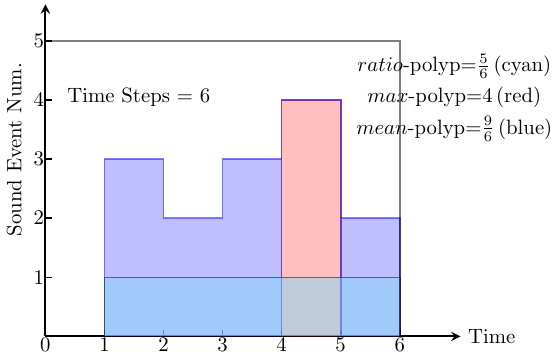}
  \captionof{figure}{\footnotesize The max polyphony level in this sound clip is 4 (show in red, time step 5), so $max$-polyp=4. The $mean$-polyp indicates the purple area, averaging them over time gets $\frac{9}{6}$. $ratio$-polyp measures polyphony\,(no fewer than two concurrent sound events) existence ratio along the time axis\,(cyan), so it is $\frac{5}{6}$.}
  \label{fig:eval_illu}
\end{figure}

Mean absolute error~(\textbf{MAE}) and mean squared error~(\textbf{MSE}) are two widely used metrics in crowd counting\,\cite{Loy2013CrowdCA, multicol_crowdcount}. Specifically, denote the ground truth count and predicted count by $y_i$ and $\hat{y_i}$ respectively, for the $i$-th sound clip. MAE is defined as ${\rm MAE}=\frac{1}{N}\sum_{i=1}^N |y_i - \hat{y_i}|$, MSE is defined as ${\rm MSE}=\sqrt{\frac{1}{N}\sum_{i=1}^N (y_i - \hat{y_i})^2}$. We also involve accuracy rate\,(\textbf{AccuRate}) to show the ratio of accurately predicted count. We introduce a tolerance term $p$, where $p=0$ means the predicted count number has to be exactly the same with ground truth number in order to be treated as an accurate counting; $p=1$ relaxes the constraint so there can be one count mismatch for an accurate counting.

The aforementioned three general metrics do not reflect the impact of sound scene nature on algorithms. We introduce three polyphony-aware metrics to quantify the sound counting difficulty level reflected by the sound scene nature. The three metrics are time-window invariant so they can be used as general metrics to quantify difficulty level of sound scene of various lengths.

\textbf{Polyphony Ratio}~($ratio$-polyp) describes the ratio of polyphony\,(at least two sound events happen at the same time) over a period of time. It binarizes each time step as either polyphonic or non-polyphonic~(monophoinc or silent) so the value lies between $[0,1]$. %It helps us to gain a rough idea of polyphony temporal coverage.

\textbf{Maximum Polyphony}~($max$-polyp) focuses on the maximum polyphony level over a time period. It is motivated by the fact that human's capability in discriminating different sound events reduces seriously when the number of temporal-overlapping sound event number increases. It is a positive integer and helps us to understand an algorithm's capability in tackling polyphony peak.

\textbf{Mean Polyphony}~($mean$-polyp) instead focuses on the averaging level of polyphony involved within a time period. It is designed to reflect an algorithm's capability in tackling the average polyphony level over an arbitrary time window.

Given $T_n$ time steps sound vector $[p_1, p_2, \cdots, p_{T_n}]$, where $p_i\geq 0$ is the sound event number happening at time step $T_i$. The three metrics are defined as,

\vspace{-4mm}
\begin{equation}
\begin{split}
    ratio{\rm-polyp} &= \frac{\sum_{i=1}^n \mathds{1}_2(p_i)}{n}; max{\rm-polyp} = \max\limits_{i=1,\cdots,n}p_i; \\
    mean{\rm-polyp}&=\frac{\sum_{i=1}^n \max(p_i-1,0)}{T_n}\\
\end{split}
\end{equation}

where $\mathds{1}_2(p_i)$ is an indicator function, it is 1 if $p_i\geq 2$, otherwise 0. With the three metrics, we can report the general metrics~(MAE, MSE) against various difficulty levels.

\section{Experiment}

We run experiments on five main category sound datasets that we commonly hear in everyday life.

1. \textbf{Bioacoustic Sound}. We focus on bird sound as bird sound is ubiquitous in most terrestrial environments with distinctive vocal acoustic properties. Specifically, we test three datasets: one real-world NorthEastUS~\cite{soundscape_dataset} dataset and other two synthesized datasets: Polyphony4Birds~(for heterophony test) and Polyphony1Bird~(for homophony test). NorthEastUS data is recorded in nature reserve in northeastern United States. It encompasses 385~minutes of dawn chorus recordings collected in July 2018, with a total of 48 bird species. The average bird sound temporal length is very short~(less than 1s) and the polyphony level~($max$-polyp and $mean$-polyp) is small. To test performance under highly polyphonic situations, we synthesize two bird sound datasets. Specifically, The first dataset contains four sounds: junco, American redhead, eagle, and rooster from copyright-free website \texttt{findsounds.com}. We call it Polyphony4Birds~(heterophony test). The second dataset contains one sound: rooster. We call it Polyphony1Bird~(homophony test). 

2. \textbf{Indoor Sound}. We count telephone ring sound, the telephone ring seed sound comes from the same copyright-free website. We follow Polyphony1Bird synthesis procedure except that the room size is much smaller~($10m \times 10m \times 3m$) to reflect indoor reverberation effect.

3. \textbf{Outdoor Sound}. We count car engine, as it is widely heard in outdoor scenario. The car engine seed sound comes from the same copyright-free website. We follow Polyphony1Bird synthesis procedure to create the dataset.

4. \textbf{AudioSet} AudioSet~\cite{audioset_data} is a large temporally-strong labelled dataset with a wide range of sound event classes, including music, speech and water. The AudioSet data tests all methods' capability in counting under large different event classes scenario. Specifically, we train model on the train dataset which has 103,463 audio clips and 934,821 labels, and test the model on the evaluation which has 16,996 audio clips and 139,538 labels. In total there are 456 sound event categories.

5. \textbf{Music Sound}. We use OpenMic2018 dataset~\cite{openmic_data} to count musical instruments. We put all discussion/results of this data to Appendix due to space limit. The direct comparison between these datasets is given in Appendix Table~1. We highly refer to Appendix. Sec.~2 for detailed discussion about the data synthesis process.

\begin{table*}[t]
\centering
\small
\caption{MSE~($\downarrow$) and MAE~($\downarrow$) on the five main category sound (six in total) datasets. The Accuracy Rate is in Appendix.}
\begin{tabular}{l|p{0.28cm}<{\centering}p{0.60cm}<{\centering}|p{0.60cm}<{\centering}p{0.51cm}<{\centering}|p{0.51cm}<{\centering}p{0.51cm}<{\centering}|p{0.51cm}<{\centering}p{0.51cm}<{\centering}|p{0.51cm}<{\centering}p{0.51cm}<{\centering}|p{0.51cm}<{\centering}p{0.51cm}<{\centering}}
\hline
\multirow{2}{*}{Method} & \multicolumn{2}{c|}{AudioSet} &  \multicolumn{2}{c|}{NorthEastUS} & \multicolumn{2}{c|}{Polyp4Birds} & \multicolumn{2}{c|}{Polyp1Bird} & \multicolumn{2}{c|}{TelepRing} & \multicolumn{2}{c}{CarEngine} \\
\cline{2-13}
 & MSE & MAE & MSE & MAE & MSE& MAE & MSE& MAE& MSE& MAE& MSE& MAE\\
\hline
Librosa-onset & 26.9 & 4.80 & 2.31 & 1.65 & 28.3 & 4.09 & 37.63 & 5.5 & 30.03 & 4.50 & 33.13 & 4.51\\
Aubio-onset & 8.50 & 1.98 & 4.91 & 1.74 & 8.43 & 1.91 & 35.33 & 5.27 & 33.20 & 4.22 & 35.13 & 4.76 \\
\hline
SELDNet~\cite{seld_dcase19} & 0.93 & 1.28 & 1.35 & 1.79 & 0.92 & 1.41 & 0.89 & 1.19 & 0.97 & 1.30 & 0.92 & 1.23\\
CRNNNet~\cite{CRNNNet} & 0.92 & 1.07 & 1.33 & 1.77 & 0.74 & 1.10 & 0.87 & 1.16& 0.92 & 1.31 & 0.86 & 1.15 \\
DND-SED~\cite{dilatedconvolution2020} & 1.10 & 1.22 & 1.19 & 1.64 & 0.95 & 1.34 & 1.04 & 1.27& 1.23 & 1.34 & 1.00 & 1.21\\
\hline
DPTasNet~\cite{multaker_ASR} & 0.81 & 1.02 & 1.11 & 1.60 & 0.97 & 1.47 & 1.22 & 1.43& 1.47 & 1.21 & 0.89 & 1.11\\
\hline
Ours DyDecNet & \textbf{0.32} & \textbf{0.73} & \textbf{1.01} & \textbf{1.19} & \textbf{0.46} & \textbf{0.92} & \textbf{0.54} & \textbf{0.85}& \textbf{0.58} & \textbf{0.89} & \textbf{0.54} & \textbf{0.87}\\
\hline
\end{tabular}
\label{table:all_MSE_MAE}
\end{table*}

\textbf{Comparing Methods}: We compare DyDecNet with three main method categories: 1) traditional signal processing methods: Librosa-onset and Aubio-onset; 2) three SED-based methods. 3) one sound source separation method. \textbf{Librosa-onset}~\cite{mcfee2015librosa} provides an onset/offset detection method for music note detection. It measures the uplift or shift of spectral energy to decide the starting time of a note. We use its onset/offset detection ability to count sound events. \textbf{Aubio-onset}~\cite{aubio2006} achieves pitch tracking by aligning period and phase. We use its pitch tracking to count.

SED-based methods build on traditional fixed TF representation, such as short time Fourier transform~(STFT) and LogMel. The TF representation is treated as a 2D image to be processed by a sequence of 2D Conv. operators. GRU~\cite{GRU} and LSTM~\cite{lstm} are often adopted to model temporal dependency. We compare three typical SED methods: 1) \textbf{CRNNNet}~\cite{CRNNNet} consists of 2D Conv. to learn multiple compressed TF representations from the input TF map. Then it concatenates them together along the frequency dimension and further feeds it to LSTM~\cite{lstm} to learn framewise representation. 2) \textbf{DND-SED}~\cite{dilatedconvolution2020} instead adopts depthwise 2D convolution and dilated convolution to avoid using RNN. 3) \textbf{SELDNet}~\cite{seldnet} is originally used for joint sound event detection and localization. It adopts 2D Conv. to convolve the 2D TF map, and bidirectional GRU to model temporal dependency. The three comparing methods' network architectures are slightly adjusted to fit our dataset. For sound source separation method, we adopt \textbf{DPTasNet}~\cite{multaker_ASR}, in which it trains a Dual-Path RNN (DPRNN) and TasNet to jointly separate each sound event and further count the event number. In this case, we treat each sound event as independent sound sources. %We call our method \textbf{Dy}adic \textbf{Dec}omposition \textbf{Net}work~(DyDecNet).

\textbf{Implementation Detail} For all datasets, all input audios are segmented into 5 second long clips, with sampling rate 24~kHz. So the input waveform has 120,000 data points and is normalized into $[-1,1]$. We train the models with Pytorch~\cite{pytorch} on TITAN RTX GPU. Network architecture of DyDecNet is given in Appendix. To train the neural network, we adopt Adam optimizer~\cite{adam_optimizer} with an initial learning rate 0.001 which decays every 20 epochs with a decaying rate 0.5. Overall, we train 60 epochs. We train each method 10 times independently and report the mean value and standard deviation. We do not report the standard deviation explicitly in the table because we find them very small~(about 0.03). We first train the comparing SED methods with both their suggested training strategy and our training strategy, then choose the one with the better performance as the final result. For the energy gain normalization we initialize them as $\alpha=0.96$, $\delta=2.$, $\gamma=0.5$, $\sigma=0.5$. The batchsize is 128.

\subsection{Experimental Result}

\begin{table*}[t]
  \begin{minipage}[t]{0.30\textwidth}
    \scriptsize
    \centering
    \caption{\footnotesize Ablation study on dyadic decomposition efficiency discussion: we compare existing methods with and without dyadic decomposition frontend.}
    \begin{tabular}{lp{0.5cm}<{\centering}p{0.5cm}<{\centering}}
    \toprule
    Method & MSE$\downarrow$ & MAE$\downarrow$ \\
    \midrule
    SELDNet~\citeyear{seld_dcase19} & 1.35 & 1.79 \\
    SELDNet\_Dydec & \textbf{1.05} & \textbf{1.43} \\
    \midrule
    CRNNNet~\citeyear{CRNNNet} & 1.33 & 1.77 \\
    CRNNNet\_Dydec & \textbf{1.20} & \textbf{1.51} \\
    \midrule
    DND-SED~\citeyear{dilatedconvolution2020} & 1.19 & 1.64 \\
    DND-SED\_Dydec & \textbf{0.89} & \textbf{1.40} \\
    \bottomrule
    \end{tabular}
    \label{table:withDyDecHead}
  \end{minipage}%
  \hfill
   \begin{minipage}[t]{0.30\textwidth}
    \scriptsize
    \centering
    \caption{\footnotesize Ablation study on traditional T-F feature for counting task: DyDecNet's dyadic decompostion frontend is replaced by various classic T-F features extractors, such STFT, LogMel, MFCC and Gabor.}
    \begin{tabular}{lp{0.5cm}<{\centering}p{0.5cm}<{\centering}}
    \toprule
    Method & MSE$\downarrow$ & MAE$\downarrow$ \\
    \midrule
    DyDecNet\_STFT & 1.35 & 1.51 \\
    DyDecNet\_LogMel & 1.33 & 1.50 \\
    DyDecNet\_MFCC & 1.32 & 1.49  \\
    DyDecNet\_Gabor & 1.33 & 1.48  \\
    \midrule
    DyDecNet & \textbf{0.85} & \textbf{1.19}\\
    \bottomrule
    \end{tabular}
    \label{table:DyDecwithTF}
  \end{minipage}%
  \hfill
     \begin{minipage}[t]{0.30\textwidth}
    %\begin{table}[t]{\textwidth}
        \tiny
        % \scriptsize
        \centering
        \captionof{table}{\footnotesize Various DyDecNet variants.}
        \begin{tabular}{lp{0.5cm}<{\centering}p{0.5cm}<{\centering}}
        \toprule
        Method & MSE$\downarrow$ & MAE$\downarrow$ \\
        \midrule
        DyDecNet\_SingScale & 1.22 & 1.43 \\
        DyDecNet\_BN & 1.07 & 1.25 \\
        DyDecNet\_noNorm & 1.15 & 1.37 \\
        \midrule
        DyDecNet & \textbf{0.85} & \textbf{1.19}\\
        \bottomrule
        \end{tabular}
        \label{table:DyDecvariant}
        \tiny
        \centering
        \captionof{table}{\footnotesize Various Counting.}
        \begin{tabular}{lp{0.5cm}<{\centering}p{0.5cm}<{\centering}}
        \toprule
        Method & MSE$\downarrow$ & MAE$\downarrow$ \\
        \midrule
        DyDecNet\_RegCount & 1.03 & 1.39 \\
        DyDecNet\_SED & 2.09 & 3.06 \\
        \midrule
        DyDecNet & \textbf{0.85} & \textbf{1.19}\\
        \bottomrule
        \end{tabular}
        \label{table:countmethod}
  \end{minipage}
  \vspace{-2mm}
\end{table*}

The quantitative result on MSE/MAE is given in Table~\ref{table:all_MSE_MAE}, accuracy rate result in Appendix. Table~2. From the two tables we can learn that DyDecNet outperforms both classic signal processing deterministic methods, comparing SED methods and sound source separation based method by a large margin, under all acoustic scenarios. DyDecNet outperforms all comparing methods in both real-world and synthesized sound datasets. It is capable of learning powerful representation from both weak sound signals~(NorthEastUS), highly polyphonic~(synthesized datasets) and heavy spectrum-overlapping, loudness-varying sound events. Moreover, we find DPTasNet~\cite{multaker_ASR} performs worse than the three SED-based methods on the two synthesized bioacoustic datasets where high-polyphony exists, which shows source separation method is not a good counting alternative in highly polyphonic situations.

\begin{figure}
\centering
\includegraphics[width=0.45\textwidth]{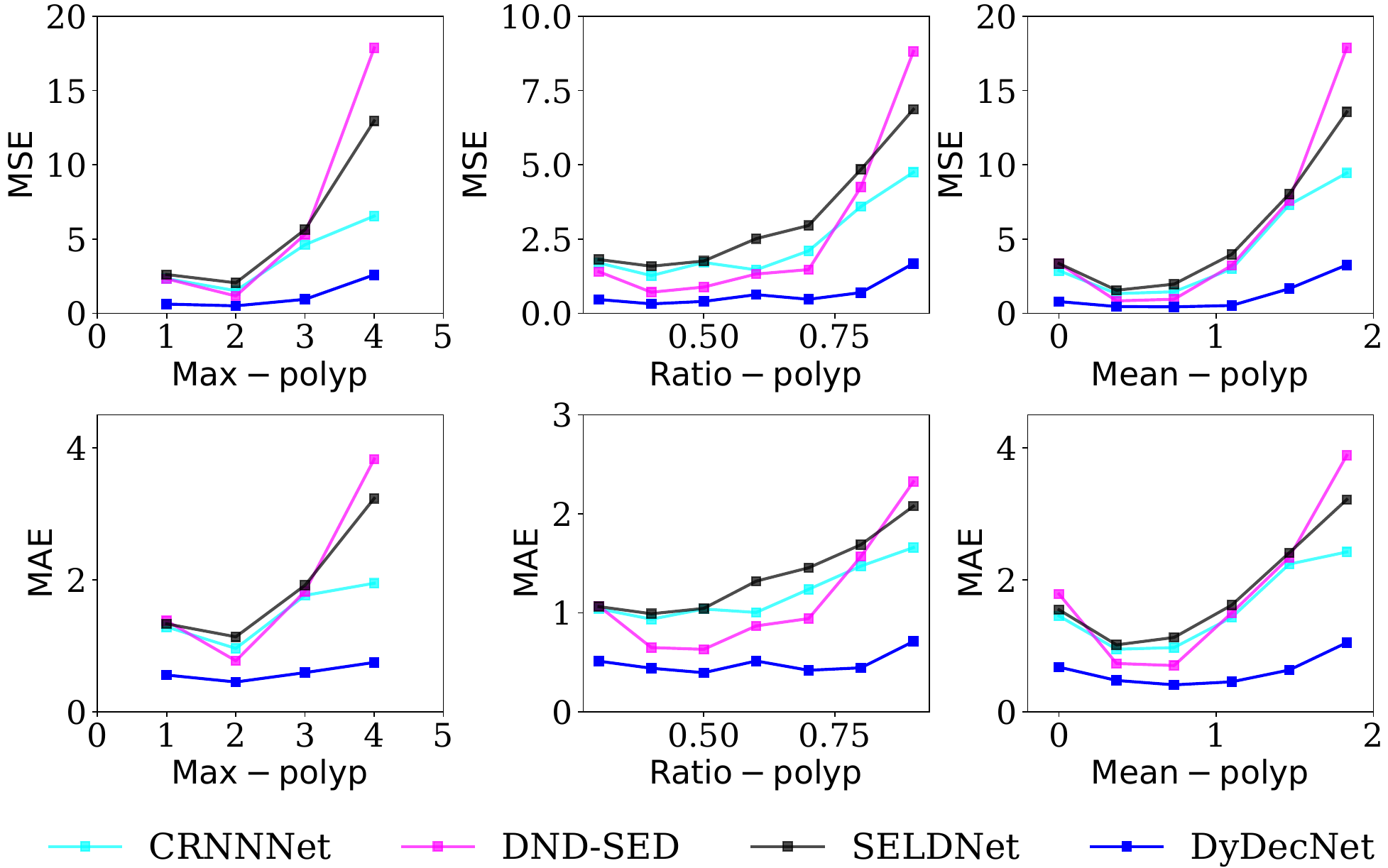}
\caption{\footnotesize MSE and MAE variation against $max$-polyp, $ratio$-polyp and $mean$-polyp on NorthEastUS dataset.}
\label{fig:enter-label}
\vspace{-2mm}
\end{figure}

At the same time, we also observe that the two signal processing deterministic methods~(Librosa-onset and Aubio-onset) generate the worst result over both SED based, source separation based methods and DyDecNet. The higher the polyphony level of the dataset, the worse performance the two deterministic methods lead to. For example, in NorthEastUS dataset with a relatively smaller polyphony level, Librosa-onset and Aubio-onset generate relatively good performance with accuracy rate~($p=1$) reaching 0.58. In our synthesized two datasets with much higher polyphony levels, however, their accuracy drops significantly to near zeros. It thus shows traditional signal processing methods do not fit for sound counting from crowded acoustic scenes.

Moreover, SED-based methods and DyDecNet produce decreasing performance from Polyphony4Birds to Polyphony1Bird and then NorthEastUS. The largest performance drop is observed on real NorthEastUS dataset, which shows counting from real-world dataset is a tough task that desires more future attention. Spectrum-overlap led by intra-class sound events is another potential challenge~(better performance on Polyp4Birds than Polyp1Bird).

The MSE/MAE variation against $max$-polyp, $ratio$-polyp and $mean$-polyp difficulty level on NorthEastUS are shown in Fig.~\ref{fig:enter-label}. We can observe that our proposed three metrics $max$-polyp, $ratio$-polyp and $mean$-polyp are effective ways to accurately quantify sound counting tasks difficulty level. The three metrics have observed dramatic performance drop as their difficulty level increases. Nevertheless, DyDecNet remains as the best one across all the three difficulty levels, showing DyDecNet outperforms the comparing methods under difficult levels discussed in this paper.

\textbf{Feature Visualization} We visualize the comparison between dyadic decomposition front-end learned TF feature and classic MFCC~\cite{MFCC} TF feature in Appendix Fig.~6. We can see DydecNet is capable of learning more discriminative feature for two temporally-overlapping and the same class sound events.

\subsection{Ablation Study}
We do ablation study on NorthEastUS data. \textbf{First}, disentangling our proposed framework's dyadic decomposition frontend and backbone network so as to figure out their individual contribution. To this end, on the one hand, we concatenate dyadic decomposition frontend to the three SED methods backbone networks so that they can learn TF representation from raw waveform. We call them SELDNet\_dydec, CRNNNet\_dydec and DND-SED\_dydec respectively. On the other hand, we feed our backbone neural network with fixed pre-extracted TF features, including short time Fourier transform\,(STFT), LogMel, MFCC and Gabor Wavelet filter. We call them DyDecNet\_STFT, DyDecNet\_LogMel and DyDecNet\_MFCC, DyDecNet\_Gabor, respectively. The results are in Table~\ref{table:withDyDecHead} and ~\ref{table:DyDecwithTF}. We can observe that: 1) replacing traditional fixed TF feature with dyadic decomposition frontend significantly improves the performance~(Table~\ref{table:withDyDecHead}). The gain stems from two-fold: our dyadic decomposition frontend enables the network to directly learn from the raw waveform so that all frequency-selective filters are adjustable during training process. Second, the dyadic progressive decomposition enables the neural network to learn robust representation for sound counting. Similarly, a huge performance drop is observed if we let our proposed backbone neural network to learn from traditional fixed TF features~(Table~\ref{table:DyDecwithTF}). Therefore, it shows that both the dyadic decomposition frontend and backbone neural networks are important for sound counting.

\textbf{Second}, we want to figure out if the dyadic decomposition is essential for sound counting, and the importance of energy normalization block. We test three variants: our network with simply single scale decomposition which means applying all filters on the raw waveform~(DyDecNet\_SingScale) which helps validate necessity of hierarchical dyadically decomposition framework; replacing Energy-normalization module with traditional batch normalization~\cite{batchnorm}~(DyDecNet\_BN); without any normalization~(DyDecNet\_noNorm). The result is in Table~\ref{table:DyDecvariant}, from which we can clearly observe that either removing energy normalization or replacing it with batch normalization significantly reduces the performance. It thus shows the importance of energy normalization.

\textbf{Lastly} We run two ablation studies to directly regress the count number and follow SED pipeline, respectively. From the result in Table~\ref{table:countmethod}, we can conclude that directly regressing sound event count number leads to inferior performance than estimating density map. Treating it as a SED problem leads to the worst performance.

\vspace{-2mm}
\subsection{Dyadic Decomposition Frontend on SELD Task}

\begin{table}
\scriptsize
	\centering
        \caption{Dyadic Frontend on SELD Task.}
    \begin{tabular}{c|cccc}
    \hline
    Method & ER~($\downarrow$) & F~($\uparrow$) &LE~($\downarrow$) &LR~($\uparrow$) \\
    \hline
     SELDNet~\citeyear{seldnet} & 0.63 & 0.46 & 23.1 & 0.69 \\
     SELDNet\_DyDec& \textbf{0.60} & \textbf{0.49} & \textbf{22.7} & \textbf{0.73} \\
     \hline
     EIN~\citeyear{ein_v2} &	0.25 &	0.82 &	8.0 & 0.86 \\
    EIN\_DyDec & \textbf{0.21} & \textbf{0.86} & \textbf{7.4} & \textbf{0.88}\\
    \hline
    SoundDet~\citeyear{sounddet} &	0.25 &	0.81 &	8.3 & 	0.82 \\
SoundDet\_DyDec	& \textbf{0.21} & \textbf{0.88} & \textbf{7.2} & \textbf{0.86} \\
     \hline
     SoundDoA~\citeyear{sounddoa} &	0.23 &	0.85 &	7.9 &	0.87 \\
SoundDoA\_DyDec & \textbf{0.20} & \textbf{0.89} & \textbf{7.4} & \textbf{0.89} \\
\hline
    \end{tabular}
    \label{tab:seld_rst}
    \vspace{-3mm}
\end{table}

To show the dyadic decomposition front-end is a general TF feature extractor, we test it on sound event detection and localisation task~(SELD). The dataset we use is TAU-NIGENS~\cite{dcase2020}, and we compare with four main methods: SELDNet~\cite{seldnet}, EIN~\cite{ein_v2} that use classic TF feature, SoundDet~\cite{sounddet} and SoundDoA~\cite{sounddoa} use learnable TF-feature. We replace their time frequency (TF) extraction front-end with dyadic decomposition network front-end to see the performance change. The result is given in Table~\ref{tab:seld_rst}, we can see that dyadic decomposition front-end exhibits generalization strength to help tackle other acoustic tasks.

% \bibliography{aaai24}

\newpage
\section{Appendix}

\section{Related Work}
\label{related_work}
Crowd counting from images or audio-visual has been thoroughly studied in recent years~\cite{multicol_crowdcount,ambientSound}, the target of which is to estimate the instance number from very crowded scenes~(e.g. pedestrian in train station) that cannot be efficiently handled by object detection methods. These methods approaching image crowd counting chronically evolve from the early detection-based~\cite{pedestrian_crowdscene} to the later regression-based~\cite{privacy_preserving_cc} and density map estimation~\cite{learn2count} methods. Accompanying these methods, various neural network architectures have been designed to achieve higher performance.

The counterpart task purely in sound, however, has been nearly ignored. Existing research mainly focus on sound event detection, including spatio-temporal sound event detection~(SELD)~\cite{sounddet,sounddoa,seldnet,ein_v2,soundsynp} from a microphone array and temporal sound event detection~\cite{gmmhmm2006,gmmhmm2010} and high-frequency time series analysis~\cite{pnas_count}. They often combine convolutional neural networks~(CNN)~\cite{cnnsed2016} and reccurrent neural network~\cite{seldnet} to separate sound sources. The datasets they work on are relatively simple, in which the sound scenes are relatively simple and contain few overlapping sound events.

The common way to process raw sound waveform is to first convert the 1D waveform into 2D time-frequency representation so that sound events' frequency property and their variation along time axis are explicitly split out. Most existing methods~\cite{ein_v2,seldnet,gmmhmm2006,gmmhmm2010} adopt Fourier transform~\cite{MFCC} or Wavelet transform~\cite{mallat_wavelet} to obtain such 2D representation, in which the whole conversion process is fixed. Some recent work~\cite{sounddet,iclr_LEAF,SincNet,Acoustic_Modeling,estimate_phoneme} re-parameterize the conversion frequency-selective filters to be learnable so that the whole neural network is able to directly learn from raw sound waveform. Experimental results show enabling the neural network to learn from the raw waveform can often achieve better performance than traditional fixed conversion. These methods, however, convert the raw waveform in a one-stage manner. Our proposed dyadic decomposition neural network instead processes the raw waveform in a dyadic multi-stage manner.

\textbf{Dyadic Network} Dyadic representation idea has been initially proposed to represent the signal hierarchically~\cite{scatter_invariant,joint_tf_scattering}, in multi-scale manner. Its core idea is to construct a bank of filters~(either learnable or fixed) so that different filter extracts different feature at a certain scale or resolution. Summarizing them together leads to more comprehensive and complete analysis. Similar idea has been widely used in the computer vision community, including pyramid feature representation for object detection~\cite{feature_pyramid_od} and semantic segmentation~\cite{feature_pyradmid_landseg,feature_pyradmid_seg}.

\textbf{Sound Source Separation} There is a massive work focusing on sound source separation~\cite{multaker_ASR,sed_sep,compmen_ssep,subakan2022resourceefficient,twostep_ssep}, in which they isolate individual sound from a mixture of audios and further assign the extracted sound to its corresponding source. While sound counting can be solved by source separation methods if the target is to count source number, our proposed sound counting is source-agnostic and it counts all sound events in a sound snippet.

\section{Sound Counting Problem Definition}
\label{scc_define}

Given a mono-channel $T$ seconds raw sound waveform $x(t)$ sampled at a fixed sampling rate $F_s$, the sound recording has recorded $N$ independent sound events $E = \{E_i = (t_s, t_e)\}_{i=1}^{N}$, each single sound event freely undergoes either stationary or moving motion in the open area. The target is to design a neural network $\mathbfcal{N}$ parameterized by $\theta$ to predict sound event number $N$ from raw sound waveform $N = \mathbfcal{N}(\theta|x(t))$. In our formulation, the counting process is class-agnostic, so all sound events are treated as instances to count, regardless of their classes. 

Three challenges make it a challenging task: 1) \textbf{Large Datasize}: microphone usually records sound at a high frequency rate~(\textit{i.e.} 24\,kHz), resulting in large data size in the raw waveform. It thus requires more accessible filters with few parameters and computation cost to process the raw sound waveform. 2) \textbf{Concurrent Sound Events\,(polyphony)}: sound events freely overlap both spatially and temporally, resulting highly polyphonic sound recording. It is a tough task to separate them apart from compressed 1D waveform. 3) \textbf{Loudness Variance and Spectrum Overlap}: sound events of the same class but different spatial location have large variance in their received loudness. They also have heavy spectrum overlap in the frequency domain. The above issues make counting a tough task.

\section{More Discussion on Dataset Creation}
\label{data_motivation_illu}

\subsection{Motivation of Polyphony4Birds and Polyphony1Bird Dataset Creation}
Our motivation for synthesizing Polyphony4Birds and Polyphony1Birds is three-fold:

\begin{enumerate}[leftmargin=*]
    \item NorthEastUS dataset has as many as 48 different kind of bird categories. It helps to test various methods' capability in tackling high bird diversity challenge.
    \item Polyphony4Birds dataset contains 4 kinds of bird sounds, but in much higher polyphony level~(in terms of $ratio$-polyp, $max-$polyp and $mean$-polyp). It helps us to test various methods' capability in tackling limited bird categories but high polyphony level~(heterophony test).
    \item Polyphony1Bird dataset contains 1 bird sound class in much higher polyphony level. This dataset involves heavy spectrum-overlap~(due to the temporal inter-category bird sounds overlap), so it helps to test various methods' capability in tackling high spectrum-overlap and high-polyphony challenge~(homophony test).
\end{enumerate}

In Polyphony4Birds dataset, 4 is an arbitrary number. We experimentally find involving 4 bird sounds is representative enough for heterophony test. We note that there are some other relevant public bird sound dataset~\cite{badc2019,tutsed2009,CRNNNet}, but we find they are not suitable for our study. For example, in TUT-SED 2009 data~\cite{tutsed2009}, the polyphony-level is small and the involved bird sound usually lasts too long~(not temporally separable and countable). Similarly, the Bird Audio Detection challenge~(BAD challenge)~\cite{badc2019} contains highly-sparse bird chirps~(very small polyphony-level sound). Moreover, the two real-world bird sound datasets~\cite{tutsed2009,badc2019} do not provide bird sound start time and end time label, so they are suitable for our study. The other synthesized dataset TUT-SED Synthetic 2016~\cite{CRNNNet} also contains very limited samples of high polyphony. The direct comparison between these datasets is given in Table~\ref{table:datasets}, from which we can see our created two datasets enjoy much higher polyphony-level, making them more suitable for our sound counting task.

\begin{table*}[h]
\centering
\small
\caption{Comparison between various sound dataset, where ``n/a'' means not available.}
\label{table:datasets}
\begin{tabular}{c|c|c|c|c|c}
\hline
Dataset & Data Source & Size & Event Classes & $mean$-polyp & $max$-polyp \\
\hline
BAD Challenge~\cite{badc2019} & Natural & 23~h & 2 & n/a & 3 \\ 
TUT-SED 2009~\cite{tutsed2009} & Natural & 18.9~h & 61 & n/a & 6 \\ 
TUT-SED Synthetic 2016~\cite{CRNNNet} & Synthetic & 9.3~h & 16 & 0.659 & 5 \\ 
\hline
NorthEastUS~\cite{soundscape_dataset} & Natural & 6.41~h & 48 & 0.1 & 4 \\
\hline
Ours Polyphony4Birds & Synthetic & \textbf{55.56~h} & 4 & \textbf{1.244} & \textbf{9}\\
Ours Polyphony1Bird & Synthetic & \textbf{55.56~h} & 1 & \textbf{1.975} & \textbf{9}\\ \hline
\end{tabular}
\end{table*}

\begin{table*}[t]
\small
\centering
\caption{Comparison between the six experiment datasets.}
\label{table:data_compare}
\vspace{-3mm}
\begin{tabular}{c|c|c|c|c|c|c}
\toprule
Scene & Dataset Name & Data Source & Size & Classes & $mean$-polyp & $max$-polyp \\
\hline
\multirow{3}{*}{Bioacoustic}&NorthEastUS~(\citeyear{soundscape_dataset}) & Natural & 6.41~h & 48 & 0.1 & 4 \\
\cline{2-7}
&Polyphony4Birds & Synthetic & \textbf{55.56~h} & 4 & \textbf{1.244} & \textbf{9}\\
&Polyphony1Bird & Synthetic & \textbf{55.56~h} & 1 & \textbf{1.975} & \textbf{9}\\
\hline
Indoor & Telephone Ring & Synthetic & 55.6~h & 1 & \textbf{1.975} & \textbf{9}\\
\hline
Outdoor & Car Engine & Synthetic & 55.6~h & 1 & \textbf{1.975} & \textbf{9}\\
\hline
Music & OpenMic~(\citeyear{openmic_data}) & Nature & 55.6~h & 20 & n/a & n/a \\ 
\bottomrule
\end{tabular}
\end{table*}

\begin{table*}[h]
\centering
\small
        \caption{Accuracy Rate results on the six datasets.}
        \begin{tabular}{l|p{0.51cm}<{\centering}p{0.51cm}<{\centering}|p{0.51cm}<{\centering}p{0.51cm}<{\centering}|p{0.51cm}<{\centering}p{0.51cm}<{\centering}|p{0.51cm}<{\centering}p{0.51cm}<{\centering}|p{0.51cm}<{\centering}p{0.51cm}<{\centering}|p{0.51cm}<{\centering}p{0.51cm}<{\centering}}
        \hline
        \multirow{3}{*}{\diagbox{Method}{Dataset}} & \multicolumn{2}{c|}{OpenMIC} &  \multicolumn{2}{c|}{NorthEastUS} & \multicolumn{2}{c|}{Polyphony4Birds} & \multicolumn{2}{c|}{Polyphony1Bird} & \multicolumn{2}{c|}{TelephoneRing} & \multicolumn{2}{c}{CarEngine} \\
        \cline{2-13}
        & \multicolumn{2}{c|}{Accu. Rate} & \multicolumn{2}{c|}{Accu. Rate} & \multicolumn{2}{c|}{Accu. Rate} & \multicolumn{2}{c|}{Accu. Rate} & \multicolumn{2}{c|}{Accu. Rate} & \multicolumn{2}{c}{Accu. Rate} \\
         & p = 0 & p = 1 & p = 0 & p = 1 & p = 0 & p = 1 & p = 0 & p = 1 & p = 0 & p = 1& p = 0 & p = 1\\
        \hline
        Librosa-onset & 0.05 &  0.03 & 0.20 &  0.58 & 0.01 &  0.02 & 0.01 &  0.02 & 0.01 &  0.01 & 0.01 &  0.01\\
        Aubio-onset & 0.09 & 0.22 & 0.12 & 0.32& 0.08 & 0.21 & 0.01 & 0.09 & 0.01 & 0.02 & 0.01 & 0.08 \\
        \hline
        SELDNet~(\citeyear{seld_dcase19}) & 0.49 & 0.75 & 0.25 & 0.62 & 0.48 & 0.71 & 0.33 & 0.81& 0.28 & 0.69 & 0.31 & 0.78\\
        CRNNNet~(\citeyear{CRNNNet}) & 0.49 & 0.85 & 0.26 & 0.64  & 0.47 & 0.82 & 0.35 &  0.81 & 0.27 &  0.76 & 0.30 &  0.80 \\
        DND-SED~(\citeyear{dilatedconvolution2020}) & 0.47 & 0.70 & 0.28 & 0.67 & 0.43 & 0.68 & 0.24 &  0.71& 0.18 &  0.63 & 0.20 &  0.68\\
        \hline
        DyDecNet & \textbf{0.68} & \textbf{0.92} & \textbf{0.40} & \textbf{0.82}  & \textbf{0.70} & \textbf{0.88} & \textbf{0.55} & \textbf{0.92} & \textbf{0.53} & \textbf{0.89}& \textbf{0.57} & \textbf{0.94}\\
        \hline
        \end{tabular}
        \label{table:all_accurate}
\end{table*}

\subsection{How to Simulate Open Area Environment}
We collect 4 seed sounds from copyright-free website~\footnote{see~\url{https://www.findsounds.com/}}: junco, American redhead, eagle, and rooster. To maximally reflect outdoor scenario, we simulate a large openarea environment $[100m, 100m, 100m]$ with one microphone at $[50m, 50m, 1m]$. The wall is associated with high sound absorption coefficient, so the reverberation is negligible so as to resemble outdoor open area scenario. We introduce a random SNR~(Signal-to-Noise Ratio) at two Gaussian means ($-33$ decibels and $-20$ decibels) at the microphone receiver. We put each seed sound at a random 3D spatial location and a random start time to imitate natural bird sounds that emit sound from a random location and random start time. A post-processing step is added to keep dataset balance between various polyphony-level metrics. 

\section{Learned Feature Visualization}
\label{feat_vis_appendix}

\begin{figure*}[h]
    \centering
    \includegraphics[width=0.80\linewidth]{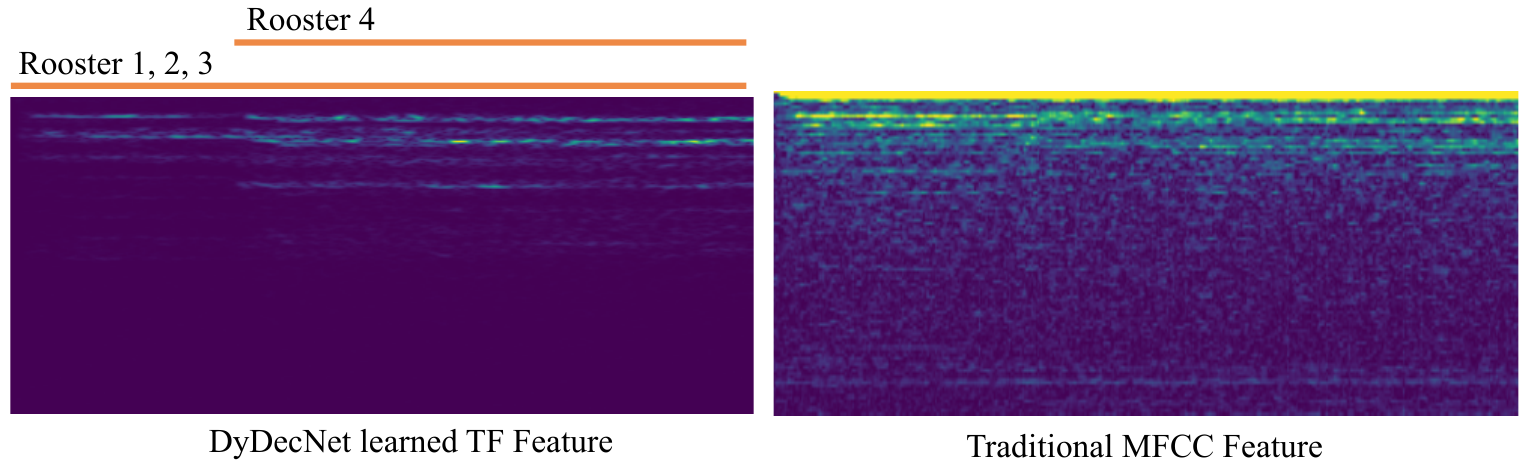}
    \vspace{-2mm}
    \caption{The comparison between DyDecNet learned TF-feature and traditional MFCC coded feature. The vertical axis indicates the frequency dimension, and the horizontal axis indicates the time axis. The orange bar on the top indicates the rooster sound start time and end time.}
    \vspace{-3mm}
    \label{fig:feat_vis_appendix}
\end{figure*}

\textbf{Feature Visualization} We visualize the DyDecNet learned time-frequency feature~(TF feature) and traditional MFCC~\cite{MFCC} on one minute long sound waveform which encodes 4 temporally overlapping sounds~(from Polyphony1Bird dataset, the 4 sounds are rooster sounds). The result is shown in Fig.~\ref{fig:feat_vis_appendix}. From this figure, we can observe that DyDecNet successfully learns frequency-separable TF representation for inter-class temporally-overlapping sounds, while traditional TF features~(in our case, MFCC) encode cluttered and mixed TF representation that is much less visually separable.

\section{More Discussion on Comparing Methods}
More detailed comparison between various methods is given in table~\ref{table:method_compare}. We can see that our proposed DyDecNet is lightweight and directly learns from sound raw waveform~(so it is end-to-end trainable). It thus strikes a good balance between model performance and model efficiency~(inference time).

\begin{table*}[t]
    \small
    \centering
    \caption{Comparison of Various Methods. The network block column labels are: 1. 1D Conv, 2. 2D Conv, 3. GRU, 4. LSTM, 5. Depthwise Conv, 6. Dilated Conv. 7. FC, 8. Bi-LSTM, 9. Bi-GRU. The inference time in tested on Intel(R) Core(TM) i9-7920X CPU, we report the average time of 100 independent tests with one 5s waveform.}
    \begin{tabular}{|c|c|c|c|c|c|}
    \hline Method Name & Input & Param Size & Network Block & Inf. Time &end2end trainable? \\ 
    \hline
    Librosa-onset & Raw Waveform & - & - & 0.1s & \XSolidBrush \\
    Aubio-onset & Raw Waveform & - & - & 0.1s & \XSolidBrush\\
    \hline
  DND-SED~\cite{dilatedconvolution2020} & STFT/LogMel & 6.9 M & 2, 5, 6, 7 & 3.0 s & \XSolidBrush\\
    CRNNNet~\cite{CRNNNet} & STFT & 4.1 M & 2, 3, 8  & 3.3 s & \XSolidBrush \\
    SELDNet~\cite{seld_dcase19} & LogMel & 0.8 M & 2, 3, 7, 9 & 1.2s & \XSolidBrush \\
    % & dnd-SED & LogMel & 6.5 M & 0.42 &  3.0 & 0.26 \\
    \hline
    DyDecNet & Raw Waveform & 3.9 M & 1, 7 & 2.7 s & \CheckmarkBold\\
    \hline
    \end{tabular}
\label{table:method_compare}
\end{table*}

\section{More Experiment Result Discussion}

\subsection{Music Dataset MAE/MSE result}
The MAE/MSE dataset result is show in Table~\cite{tab:music_rst}. From this table we can observe that our proposed DyDecNet is the best-performing method.

\begin{table}[t]
    \scriptsize
    \centering
    \caption{Music Dataset MSE/MAE result.}
    \begin{tabular}{c|ccccc}
    \hline
        Method & SELDNet & CRNNNet & DND-SED & DPTasNet & \textbf{DyDecNet}  \\
        \hline
        MSE/MAE &  0.90/1.37 & 0.71/1.00 & 0.93/1.27 & 0.44/0.91 & \textbf{0.32/0.72} \\
    \hline
    \end{tabular}
    \label{tab:music_rst}
\end{table}

\begin{table}[t]
    \centering
    \caption{DyDecNet with Traditional T-F feature and learnable energy normalization module}
    \begin{tabular}{lcc}
    \hline
    Method & MSE$\downarrow$ & MAE$\downarrow$ \\
    \toprule
    DyDecNet\_STFT & 1.30 & 1.46 \\
    DyDecNet\_LogMel & 1.27 & 1.47 \\
    DyDecNet\_MFCC & 1.26 & 1.44  \\
    DyDecNet\_Gabor & 1.30 & 1.43  \\
    \midrule
    DyDecNet & \textbf{0.85} & \textbf{1.19}\\
    \bottomrule
    \end{tabular}
    \label{table:withDyDecHead1}
\end{table}

\begin{table}[t]
    \centering
    \caption{DyDecNet with Traditional T-F feature}
    \begin{tabular}{lcc}
    \toprule
    Method & MSE$\downarrow$ & MAE$\downarrow$ \\
    \midrule
    DyDecNet\_STFT & 1.35 & 1.51 \\
    DyDecNet\_LogMel & 1.33 & 1.50 \\
    DyDecNet\_MFCC & 1.32 & 1.49  \\
    DyDecNet\_Gabor & 1.33 & 1.48  \\
    \midrule
    DyDecNet & \textbf{0.85} & \textbf{1.19}\\
    \bottomrule
    \end{tabular}
    \label{table:DyDecwithTF1}
\end{table}

\subsection{Experiment on NorthEastUS Dataset and Telephone Ring Dataset}

More detailed experimental result~(MAE variation) on NorthEastUS is given in Fig.~\ref{fig:eval_ne_final}, from which we can observe that with the increasing of $max$-polyp, $ratio$-polyp and $mean$-polyp, all methods~(including our DyDecNet) reduces their performances. The three comparing methods~(CRNNNet~\cite{CRNNNet},DND-SED~\cite{dilatedconvolution2020}, and SELDNet~\cite{seld_dcase19}) have observed sharp performance drop when the our proposed three sound counting difficulty levels increases, whereas our proposed DyDecNet largely mitigates the challenge caused by higher counting difficulty level~(the blue line increases slightly as the counting difficulty level increases). It thus shows 1) our proposed $max$-polyp, $ratio$-polyp and $mean$-polyp are capable of accurately measuring sound counting task difficulty level from different perspectives; 2) our proposed DyDecNet is capable of mitigating these sound counting difficulties.

\begin{figure}[t]
    \centering    \includegraphics[width=0.95\linewidth]{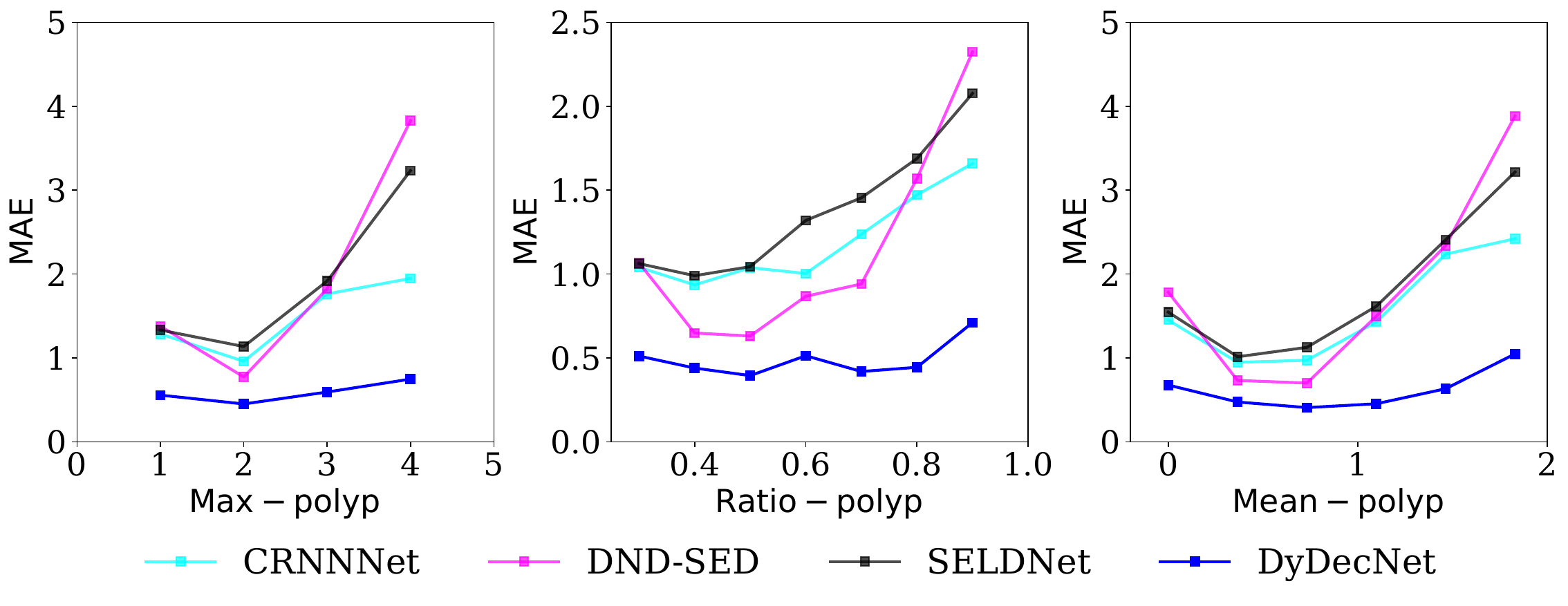}
    \vspace{-3mm}
    \caption{MAE variation against $max$-polyp, $ratio$-polyp and $mean$-polyp on NorthEastUS Dataset.}
    \label{fig:eval_ne_final_v2}
\end{figure}

\subsection{More Result on Polyphony1Bird and Polyphony4Birds Datasets}

\begin{figure}[t]
    \centering
    \includegraphics[width=0.95\linewidth]{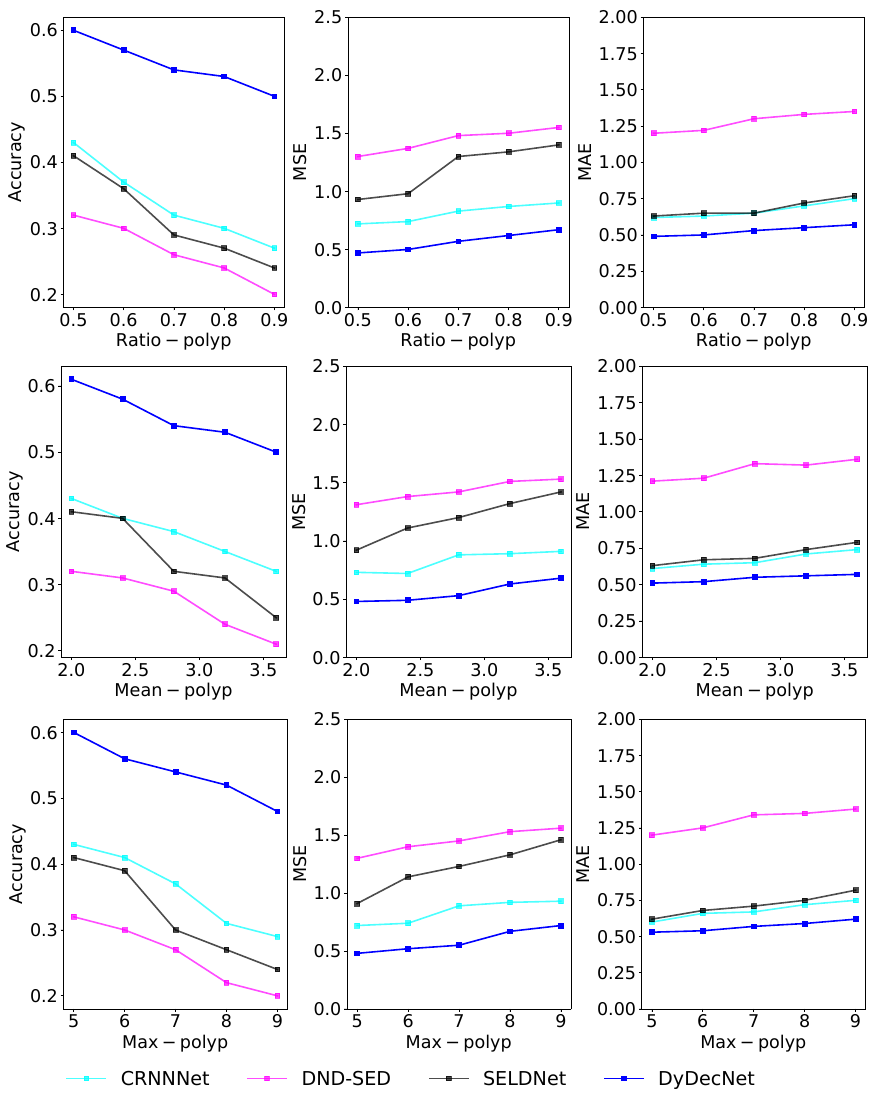}
    \vspace{-3mm}
    \caption{AccuRate/MSE/MAE variation against $max$-polyp, $ratio$-polyp and $mean$-polyp on Polyphony1Bird Dataset.}
    \label{fig:eval_1bird_final}
\end{figure}

\begin{figure}[ht]
    \centering
    \includegraphics[width=0.95\linewidth]{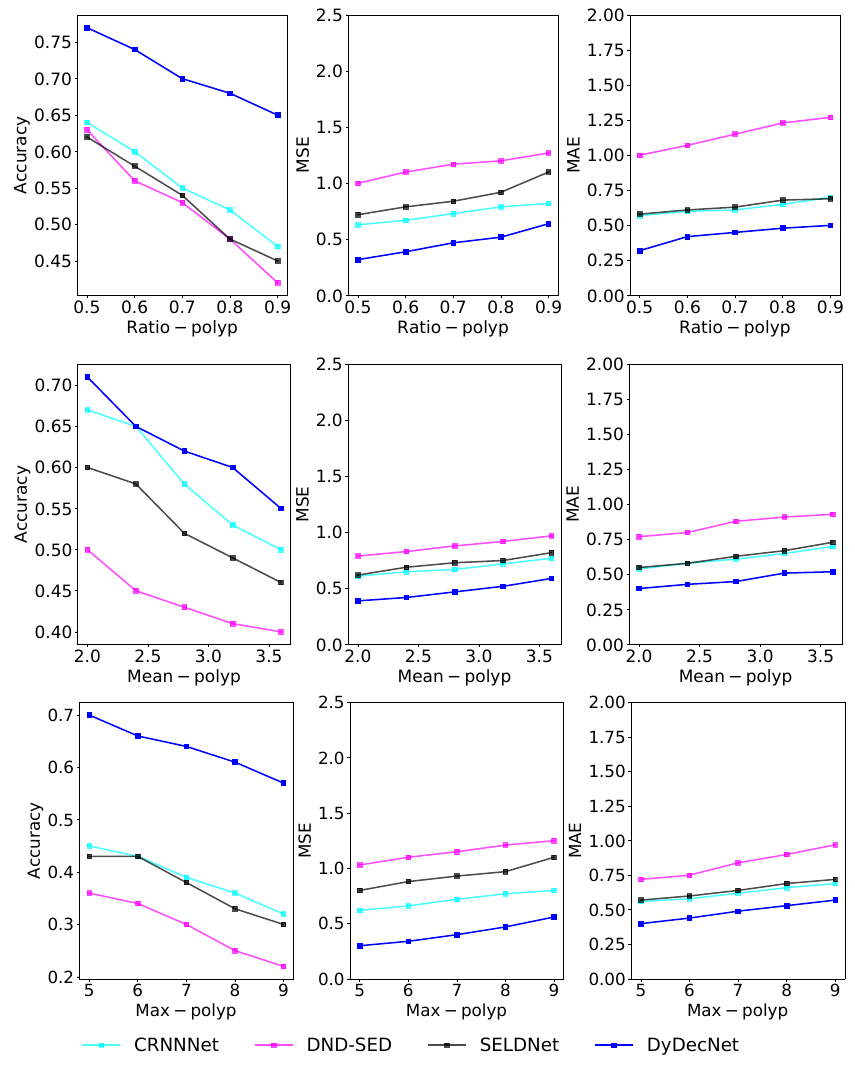}
    \caption{AccuRate/MSE/MAE against $max$-polyp, $ratio$-polyp and $mean$-polyp on Polyphony4Birds Dataset.}
    \label{fig:eval_4bird_final}
\end{figure}

We also provide the detailed results for Polyphony1Bird and Polyphony4Birds in Fig.~\ref{fig:eval_1bird_final} and Fig.~\ref{fig:eval_4bird_final}, respectively. They contain the accuracy rate, MSE and MAE variation against $max$-polyp, $ratio$-polyp and $mean$-polyp. From the two figures, we can get similar conclusion as of NorthEastUS dataset~(Fig.~\ref{fig:eval_ne_final_v2}): with the increasing of $max$-polyp, $ratio$-polyp and $mean$-polyp, all methods' performance gradually reduces. Our proposed DyDecNet stays as the best-performing one under all sound counting difficulty level metrics. Specifically, we can see that:

\begin{itemize}
    \item All methods give the best performance on Polyphony4Birds dataset, second best performance on Polyphony1Bird dataset, and the worst performance on NorthEastUS dataset. It thus shows 1) spectrum-overlap due to high inter-class sound overlap temporally~(represented by Polyphony1Bird dataset) remains as a challenge for sound counting task. 2) sound counting in open area where noise pollution, high sound diversity~(in our case, diversity means bird categories, we have 48 bird classes in NorthEastUS dataset), and small labelled data availability exist remains as another challenge for sound counting task. We hope to attract more researchers to consider sound counting task in more challenging scenario.
    \item We do not observe such sharp performance drop~(as we observed on NorthEastUS dataset) on our two synthetic datasets, which is in contrast with the real-world dataset NorthEastUS. It thus shows real-world sound counting task becomes increasingly challenging when our proposed three sound counting difficulty level metrics increase. We guess the large model and large training dataset are needed to achieve better performance, which can be treated as a future research direction. 
\end{itemize}

\begin{table*}[t]
    \centering
    \caption{Dyadic Decomposition Neural Network Architecture Illustration. Input audio is 5~s long, sampling rate is 24~k Hz. The input waveform and intermediate features are in $[channel num, height, width]$ format. In the dyadic decomposition front-end, the constructed filters' initialized trainable frequency cutoffs~(high frequency cutoff and low frequency cutoff) evenly divide the frequency range of the input sound waveform~(half of sampling rate). In our design, the filter number doubles as the ``Depth'' increases by 1. So each filter in the preceding depth associates with two child filters in the next depth, in which the two child filters carried frequency cutoffs evenly divide the frequency cutoff of their parent filter~(see Fig.~1 in the main paper). In our implementation, we just connect a filter in the preceding depth with its two child filters in the next depth. We organize filters in channel dimension. In the dyadic decomposition front-end, each filter is instantiated as $Sinc(\cdot)$ filter, which comprises of a learnable high frequency cutoff and a low learnable frequency cutoff.}
    \begin{tabular}{|c|c|c|}
        \hline
        Layer Name & Filter Num.  & Output Size \\
        \hline
        \multicolumn{3}{|c|}{\textbf{Input Size}: 5~s audio waveform: [1, 1, $24000\times 5$]}\\
        \hline
        \multicolumn{3}{|c|}{\textbf{Dyadic Decomposition Front-End}}\\
        \hline
        Dyadic Decomp. Depth 1 & filter num = $2^1$, 2x downsampple  & [2, 1, $12000\times 5$]\\
        Dyadic Decomp. Depth 2 & filter num = $2^2$, 2x downsampple & [4, 1, $6000\times 5$]\\
        Dyadic Decomp. Depth 3 & filter num = $2^3$, 2x downsampple  & [8, 1, $3000\times 5$]\\
        Dyadic Decomp. Depth 4 & filter num = $2^4$, 2x downsampple & [16, 1, $1500\times 5$]\\
        Dyadic Decomp. Depth 5 & filter num = $2^5$, 2x downsampple & [32, 1, $750\times 5$]\\
        Dyadic Decomp. Depth 6 & filter num = $2^6$ & [64, 1, $750\times 5$]\\
        Dyadic Decomp. Depth 7 & filter num = $2^7$ & [128, 1, $750\times 5$]\\
        Dyadic Decomp. Depth 8 & filter num = $2^8$ & [256, 1, $750\times 5$]\\
        \hline
        \multicolumn{3}{|c|}{\textbf{Backbone Network}} \\
        \hline
        Per-channel Pool & SincLowPass Filters, stride = 5 & [256, 1, $750\times 1$] \\
        Cross-channel Conv. & 1D Conv., filter num = 512 & [512, 1, $750\times 1$] \\
        \hline
        Per-channel Pool & SincLowPass Filters, stride = 5 & [512, 1, $150\times 1$] \\
        Cross-channel Conv. & 1D Conv., filter num = 1024 & [1024, 1, $150\times 1$] \\
        \hline
        Per-channel Pool & SincLowPass Filters, stride = 3 & [1024, 1, $50\times 1$] \\
        Cross-channel Conv. & 1D Conv., filter num = 512 & [512, 1, $50\times 1$] \\
        \hline
        Cross-channel Conv. & 1D Conv., filter num = 256 & [256, 1, $50\times 1$] \\
        FC & FC, output\_feat = 1 & [50, 1] \\
        \hline
    \end{tabular}
    \label{dydecnet_network}
\end{table*}

\subsection{Counting on More Birds Classes}
In the main paper, our two synthetic datasets Polyphony4Birds and Polyphony1Bird have just involved limited bird classes~(up to 4). We naturally want to figure out the performance of all methods~(including DyDecNet and the other three comparing methods) under more bird classes situation. We thus follow more the same data creation procedure to synthesize four extra datasets. They contain 2/6/8/10 bird classes, respectively. The extra bird seed sound classes are collected from \texttt{findsound.com} too. The quantitative result is given in table~\ref{table:multibirds}, from which we can learn that all comparing methods have observed performance increasing when the bird classes reach to 6~(values in bold font), then performance decline when bird classes increase to 8 or 10; DyDecNet reaches the best performance around 8 bird classes, then begin to decline. It thus shows: 1) all methods can successfully handle a reasonable amount of bird classes~(in our case, maximum bird classes are 8), given the model parameter size budget~(less than 10~M) discussed in this paper. When we have to handle much larger bird diverse classes, we might need much larger model~(which remains as a future research topic to figure out the relationship between model size and sound counting class diversity); 2) Our proposed DyDecNet exhibits strong capability sound counting in diverse bird classes than the three comparing methods~(it reaches the best performance at a higher bird classes~(8 bird class)).

\begin{table*}[h]
    \centering
    % \small
    \caption{MSE/MAE/AccuRate results on Multiple Bird Classes. The three values split by `/' in each entry indicate MSE, MAE and AccuRate, respectively. For the AccuRate, we just report the accuracy rate under $p=0$. Following the experiment setting in the main paper, we run each model 10 times independently. We do not report the standard deviation, they are all within 0.003.}
    \begin{tabular}{l|ccc|c}
    \toprule
    Bird Classes & SELDNet~(\citeyear{seld_dcase19})&CRNNNet~(\citeyear{CRNNNet}) & DND-SED~(\citeyear{dilatedconvolution2020}) & DyDecNet\\
    \midrule
    1~Bird & 0.89\ /\ \textbf{1.19}\ /\ 0.33 & 0.87\ /\ 1.16\ /\ 0.35 & 1.04\ /\ 1.27\ /\ 0.24 & 0.54\ /\ 0.85\ /\ 0.55\\
    4~Birds & 0.92\ /\ 1.41\ /\ 0.48 & 0.74\ /\ 1.10\ /\ 0.47 & 0.95\ /\ 1.34\ /\ 0.43 &  0.46\ /\ 0.92\ /\ 0.70 \\
    6~Birds & \textbf{0.88}\ /\ 1.40\ /\ \textbf{0.49} & \textbf{0.72}\ /\ \textbf{1.09}\ /\ \textbf{0.49} & \textbf{0.93}\ /\ \textbf{1.26}\ /\ \textbf{0.48} &  0.43\ /\ 0.90\ /\ 0.74 \\
    8~Birds & 0.93\ /\ 1.45\ /\ 0.40 & 0.75\ /\ 1.13\ /\ 0.44 & 0.95\ /\ 1.37\ /\ 0.42 &  \textbf{0.41}\ /\ 0.88\ /\ \textbf{0.77}\\
    10~Birds & 0.95\ /\ 1.49\ /\ 0.37 & 0.78\ /\ 1.17\ /\ 0.39 & 0.97\ /\ 1.40\ /\ 0.37 &  0.44\ /\ \textbf{0.82}\ /\ 0.72\\
    \bottomrule
    \end{tabular}
    \label{table:multibirds}
\end{table*}

% \subsection{Noise Discussion}

\subsection{Learnable Energy Normalization with Traditional T-F Feature}
\label{energy_normal_traditionalTF}
To test the efficiency of our proposed energy normalization module, especially combining them with traditional one-stage T-F features, we explicitly add one learnable energy normalization module just after the T-F feature extracted by traditional time frequency feature extractors. The comparison is given in Table~\ref{table:withDyDecHead1} and Table~\ref{table:DyDecwithTF1}, from which we can observe that performance of traditional T-F feature slightly increases after introducing the energy normalization module. It thus shows the necessity of energy normalization for sound counting task in high-polyphonic situation. However, they still lead to inferior performance than DyDecNet, which shows hierarchically dyadic decomposition with energy normalization is essential for sound counting task.

\section{Network Architecture}
The DyDecNet architecture is shown in Table~\ref{dydecnet_network}.

\clearpage
\bibliography{aaai24}

\end{document}